\documentclass[a4paper]{jpconf}
\usepackage{graphicx}
\usepackage{bm}
\begin{document}
\title{Impurity states and Localization in Bilayer Graphene: the Low Impurity Concentration Regime}

\author{H. P. Ojeda Collado$^1$, Gonzalo Usaj$^1$, C. A. Balseiro$^1$ }
\address{$^1$ Centro At{\'{o}}mico Bariloche and Instituto Balseiro,
Comisi\'on Nacional de Energ\'{\i}a At\'omica, 8400 Bariloche, and CONICET,
Argentina}

\ead{hpablo1988@gmail.com}

\begin{abstract}
We study the problem of non-magnetic impurities adsorbed on bilayer graphene in the diluted regime. We analyze the impurity spectral densities for various concentrations and gate fields. We also analyze the effect of the adsorbate on the local density of states (LDOS) of the different C atoms in the structure and present some evidence of strong localization for the electronic states with energies close to the Dirac point.
\end{abstract}


The problem of adatoms in graphene has been the subject of an intense activity for they could be used to modify and control the electronic properties of the material. Diluted adatoms or molecules necessarily generates disorder  \cite{DasSarma2011} and in some regimes may lead to strong localization of the electronic states \cite{Ostrovsky2006,Gattenloehner2013,Roche2012,Hong2011,Usaj2014,Cresti2013}. Electron localization in graphene is quite peculiar, Dirac fermions tend to elude localization in systems with Anderson-type disorder. However impurities leading to short range disorder at the atomic scale generate inter-valley mixing and break the symplectic symmetry opening the route to strong localization. The problem of adatoms and electron localization in bilayer graphene (BLG), although considered by several groups \cite{McC0, Nilsson}, received much less attention \cite{Ka, GCN}. In the most common structure of BLG, known as the Bernal stacking, only one of the two non-equivalent sites $(A,B)$ of the honeycomb lattice of the top layer lies on top of a site of the bottom layer. The resulting structure, shown in Fig. 1, induces a weak coupling of the two layers. The unit cell has four carbon atoms leading to four bands, two of them having a parabolic dispersion relation around the $K$ and $K^\prime$ points of the Brillouin Zone, touch each other at the Fermi energy. 

In most of the experimental setups, BLG lies on top of a substrate and the impurities are adsorbed on the top layer only. Due to the difference of the $A$ and $B$ sites of the layer, there is a small difference in the absorption energy on the two inequivalent sublattices. This difference favors absorption on the $B$ sites and in what follows we assume that all impurities are on the $B$ sublattice.
A very interesting aspect of BLG is its response to a gate field \cite{McC1, GCNP, McC2, Otha, Castro}. An electric field perpendicular to the layers opens a gap at the Fermi level an effect that can be used to modify the impurity states. Here we study the problem of non-magnetic impurities adsorbed on BLG in the diluted regime. We analyze the impurity spectral densities for various concentrations and gate fields.  The effect of the adsorbate on the local density of states (LDOS) of the different C atoms in the structure is analyzed and we present some evidence of strong localization for the electronic states with energies close to the Dirac point.

The Hamiltonian of the system includes the bilayer Hamiltonian $H_{BLG}$, the impurity contribution $H_{imp}$ and the hybridization term $H_{hyb}$. In what follows, as there are no spin effects we ignore the spin index.
\begin{eqnarray}\nonumber
H_{BLG}=&&-\sum_{j,\bm{k}}[V(-1)^{j}(a^{\dagger}_{j\bm{k}}a_{j\bm{k}}+b^{\dagger}_{j\bm{k}}b_{j\bm{k}})
+t(\phi({\bm{k}})\,a^{\dagger}_{j\bm{k}}b_{j\bm{k}}+\phi^{*}({\bm{k}})\,b^{\dagger}_{j\bm{k}}a_{j\bm{k}})]\\
&& -\sum_{\bm{k}}t_{\perp}(a^{\dagger}_{1\bm{k}}b_{2\bm{k}}+b^{\dagger}_{2\bm{k}}a_{1\bm{k}})
\end{eqnarray}
here $a_{j\bm{k}}$ and $b_{j\bm{k}}$ destroy electrons with wavevector $\bm{k}$ in sublattices $A$ and $B$ respectively, the subindex $j=1$ ($j=2$) refers to the top (bottom) plane. $V$ is the potential induced by the gate voltage, $t$ and $t_{\perp}$ are the intra and inter-plane hoppings, respectively and $\phi(\bm{k})=\sum_{\bm{\delta}}e^{i\bm{k}\cdot\bm{ \delta}}$ where $\{\bm{\delta}\}$ are the three vectors connecting one site with its neighbors in the same plane. We consider adatoms that are bounded to a single C atom,
$H_{imp}=\sum_{l}'\varepsilon_0 f^{\dagger}_{l}f_{l}$
with $f_{l}$ the destruction operator of an electron on the impurity orbital of the
adatom at site $l$, $\varepsilon_0$ is the energy of the orbital and the sum runs over the sites of the carbon lattice having an impurity on top. Finally
$H_{hyb}=\gamma \sum_{l}' (f^{\dagger}_{l}b_{1,l}+b^{\dagger}_{1,l}f_{l})$
here $b_{1,l}=1/\sqrt{N}\sum_{\bm{k}}e^{i{\bm{k}}\cdot{\bm{R}}_{l}}b_{1,\bm{k}}$ where ${\bm{R}}_{l}$ is the coordinate of site $l$. The parameters used to describe the BLG are $t=0.25$ eV, $t_{\perp} = t/9$ and we take $\gamma = 2t$ \cite{SF, SH}. Three values of the bias voltage $V$ are taken; $V=-0.02 t$, $V=0$eV, $V=+0.02t$ and without any loss of generality we take $\varepsilon_0 \ge 0$.

We first present results for the impurity contribution $\rho_{imp}(\omega)$ to the total density of states (DOS). We use the Chebyshev polynomials method which has proven to be very efficient to deal with realistic impurity concentrations \cite{Ka,Weisse2006,Covaci2010}. 
\begin{figure}[h!]
\begin{center}
\includegraphics[width=0.9\textwidth]{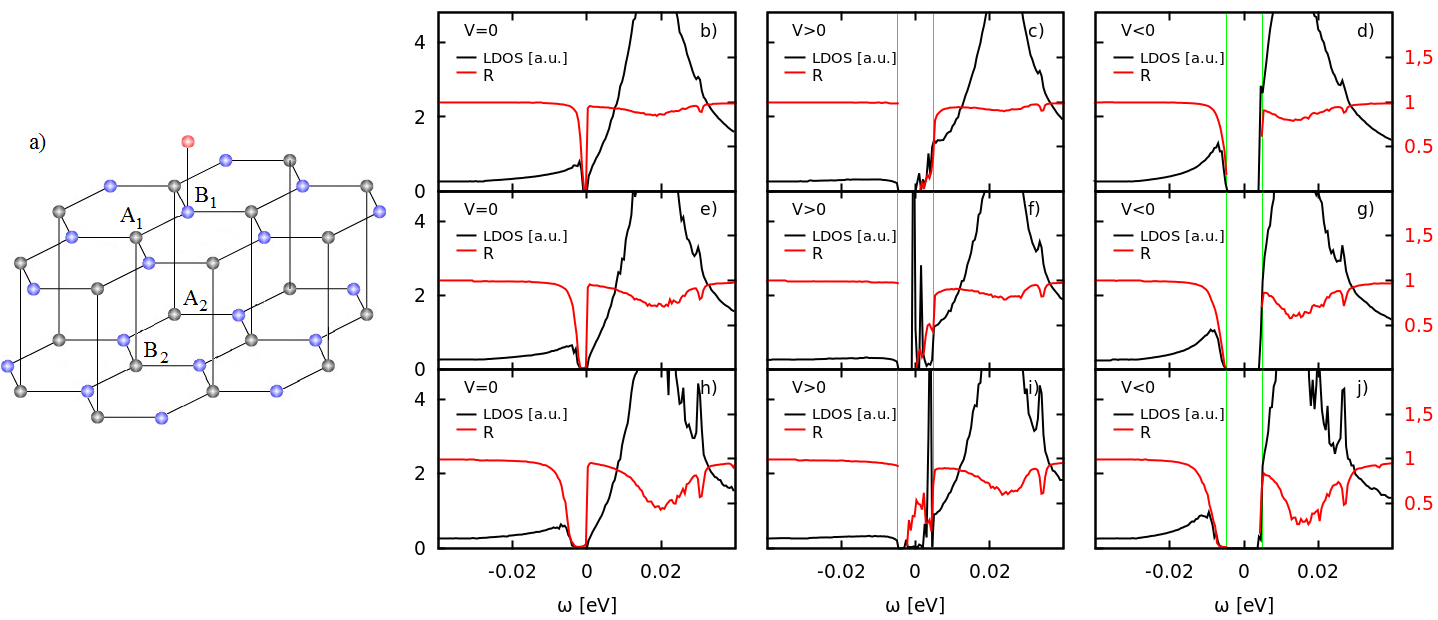}

\caption{Structure of the bilayer graphene (a) and impurity spectral densities (b-j), left scale. Upper, central and lower panels correspond to impurity concentrations $1/1000$, $1/500$ and $1/250$ respectively. The three columns correspond to unbiased BLG (left), positive (centre) and negative (right) bias. The red thick line is the function $R(\omega)$, right scale, see text. }
\label{fig1}
\end{center}
\end{figure}

The average impurity spectral density is then given by
$\rho_{imp}(\omega)=-\frac{1}{\pi}\langle\mathrm{Im}{\mathcal{G}_{ll}}\rangle_\mathrm{avg}$
where $\mathcal{G}_{ll}$ is the retarded impurity propagator and $\langle\dots\rangle_\mathrm{avg}$ indicates the configurational average over the impurities. The results for three different concentrations are shown in Fig~1. There, the three columns correspond to different values of the gate voltage $V$, the rows to different impurity concentrations. Interestingly, for $V=0$ the impurity spectral density shows a gap close to the Dirac point that increases with increasing impurity concentration. This effect is reminiscent of the gap induced in monolayer graphene when impurities are adsorbed in a single sublattice \cite{desbalance}. 
For a non-zero gate voltage $V$, the pristine BLG develops a gap at the Dirac point (indicated by vertical lines in the figure). For positive $V$ the gap is partially filled by impurity states. For large impurity concentrations the gap closes, while for small concentrations a reduced gap remains. In the thermodynamic limit and for all cases discussed above, we expect the gaps to be just pseudogaps with exponentially small DOS \cite{tail}. The case of negative $V$ is completely different: the gap of the pristine BLG remains unaltered for small or moderate values of $|V|$. This effect can be understood by looking at the response of a single impurity to the gate field \cite{mkmi}. In the single impurity case and for $ \varepsilon_0 \ne 0$ and small values of $V$  a bound state appears in the gap only for one polarity of the field.

To better understand the nature of the electronic states and the effect of the gates, we calculate the LDOS at the different C atoms, in the upper and lower layer. Results for different concentrations and polarities of the bias field are shown in Fig 2. In all cases the impurity contribution to the LDOS for small energies is large at the $A1$ sublattice, an effect that is also characteristic of impurities adsorbed on top of a single $B1$-carbon atom in monolayers \cite{SH}.  

\begin{figure}[h!]
\begin{center}
\includegraphics[width=0.9\textwidth]{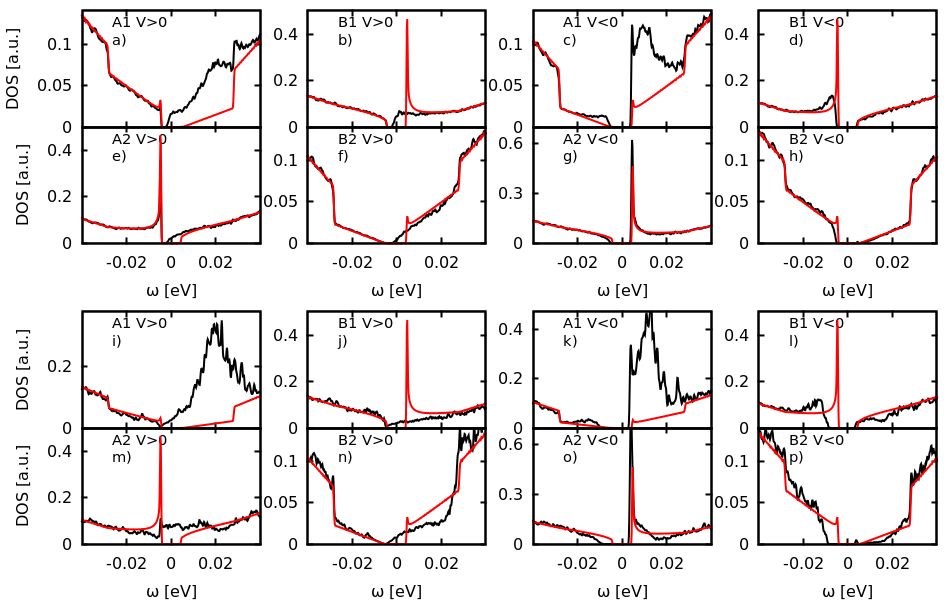}

\caption{LDOS of the four sublattices for  impurity concentrations $1/1000$ (upper panels) and $1/250$ (lower panels) and different polarities of the bias field. Smooth curves correspond to the pristine sample. Note the different scales of the panels}
\label{fig2}
\end{center}
\end{figure}
For low concentrations the LDOS of the other sublattices---namely $B1$, $A2$ and $B2$---present only small modifications. In particular for $V>0$ the valence band remains essentially unaltered with its $1D$-like van Hove singularity in the $A2$ sublattice. This strongly suggests that at least for this polarity  there is no strong localization of the electronic states in the valence band. We could draw similarly conclusions from the structure of the LDOS of the $A2$ sublattice for $V<0$ as illustrated in the upper panels of Fig 2. As the concentration increases (lower panel of the figure), the modifications of the LDOS become more important and it is necessary to look for a better and more qualitative criterion for localization. To this end we evaluate  the function $R(\omega)=\rho_{typ}(\omega)/\rho(\omega)$ where \cite{rw}
\begin{equation}
\label{Rw}
\rho_{typ}(\omega)=\left[\prod_{l=1} ^N\rho_l(\omega)\right]^\frac{1}{N}   \mbox{ and }         \quad \rho(\omega)=\frac{1}{N}\sum_{l=1} ^{N}\rho_l(\omega)\,.
\end{equation}
Here $N$ is the number of impurities and $\rho_{l}(\omega)=-\frac{1}{\pi}\mathrm{Im}{\mathcal{G}_{ll}}$ is the spectral density of the $l^{th}$ impurity. When, for a given energy $\omega$, the states are extended $\rho_{l}(\omega)$ has small fluctuations and $\rho_{typ}(\omega)\approx \rho(\omega)$ resulting in $R(\omega)\approx 1$. In contrast, if states are localized $\rho_{typ}(\omega)$ is dominated by small values of $\rho_{l}(\omega)$ and $R(\omega)\ll1$, with $R(\omega) \to 0$ as $N \to \infty$. We stress that our analysis is based only on the properties of the impurity spectral densities and gives a very qualitative estimation of the energy range where we may expect strong localization effects. 
While, as mentioned above, the Chebyshev polynomials method is very efficient to evaluate the average spectral densities, it becomes numerically costly to evaluate a large number of $\rho_{l}(\omega)$. To estimate $R(\omega)$ we then resort to the method described in Ref. \cite{Usaj2014}. The impurity propagator matrix $\bm{\mathcal{G}}$ with matrix elements $\mathcal{G}_{ij}(\omega)$ satisfies the Dyson equation
\begin{equation}
\left[(\omega+i0^+-\varepsilon _{0})\bm{I}-\gamma^{2}\tilde{\bm{g}}\right]\bm{\mathcal{G}}=\bm{I}\,,
\end{equation}
where $\bm{I}$ is the unit matrix and $\tilde{\bm{g}}$ is a matrix whose elements are
the propagators of pristine biased BLG, $g_{B1i,B1j}(\omega)$, between impurity sites adsorbed on the $B1$ sublattice. For large distances and low frequencies we evaluate $g_{B1i,B1j}(\omega)$ in the continuous limit \cite{RKKY}. The required spectral densities are the imaginary part of the diagonal terms of the matrix $\bm{\mathcal{G}}$.

Results for $R(\omega)$ are presented in Fig. 1. They show that strong localization effects are to be expected for energies very close to the Dirac point. In the case of positive bias, when the gate induced gap is filled by impurity states, these states in the gap are strongly localized. Our results suggest that, independently of the bias and for large impurity concentration, localization effects are also to be expected in the energy window close to the maximum of the impurity LDOS. This last effect is also observed in monolayer graphene although the localization length may be quite different in the two systems. Away from these energies, only weak localization effects are likely. A more quantitative estimation of the localization phenomena requires evaluation of the localization length.

As a final remark we mention that for the parameters used in the present work, that are suitable to describe fluorine on graphene, and the small bias voltages of the figures,  a single impurity on an $A1$ sites does not generate a bound state inside the gap for positive $V$, for negative $V$ there is a bound state exponentially close to the gap edge. As a consequence, a small amount of impurities added to the $A1$ sublattice would not change the results obtained for small energies (within the field induced gap). 

In summary, we have presented results for diluted impurities on BLG with and without gate voltages that open a gap in the pristine sample. We have shown the existence of drastic effects of the polarity of the electric field. For impurity parameters appropriate to describe fluorine on BLG, the gap induced in the pristine sample for positive polarity is filled with strongly localized states. Conversely, for negative polarity the gap remains. In all cases, strong localization occurs only at low energies.

We acknowledge useful discussions with J. Sofo and financial support from PICT
Bicentenario 2010-1060 from ANPCyT, PIP 11220080101821 and 11220110100832 from CONICET and 06/C400 and 06/C415 SeCyT-UNC. HPOC acknowledges scholarship from ICTP.

%

\end{document}